\documentclass[twocolumn,preprintnumbers,amsmath,amssymb,prb]{revtex4}

\usepackage{graphicx}
\usepackage{dcolumn}
\usepackage{bm}
\usepackage{amssymb}
\usepackage{epstopdf}
\usepackage{color}

\begin{document}

\title{Electron pairing in nanostructures driven by an oscillating field}
\author{O. V. Kibis}\email{Oleg.Kibis(c)nstu.ru}

\affiliation{Department of Applied and Theoretical Physics,
Novosibirsk State Technical University, Karl Marx Avenue 20,
Novosibirsk 630073, Russia}

\begin{abstract}
It is shown theoretically that the confinement of an electron at a
repulsive potential can exist in nanostructures subjected to a
strong high-frequency electromagnetic field. As a result of the
confinement, the metastable bound electron state of the repulsive
potential appears. This effect can lead, particularly, to electron
pairing in nanostructures containing conduction electrons with
different effective masses.
\end{abstract}

\maketitle

\section{Introduction}
Achievements in laser physics and microwave technique have made
possible the use of an oscillating field as a tool to control
physical properties of various systems~\cite{GoldMan_14,Bukov_15}.
Particularly, conduction electrons driven by a strong
high-frequency field are actively studied to exploit features of
composite field-matter states in various nanostructures, including
superlattices~\cite{Holthaus_92,Holthaus_95}, semiconductor
quantum wells~\cite{Kibis_12,Kibis_14}, quantum
rings~\cite{Kibis_11,Koshelev_2015},
graphene~\cite{Kibis_10,Glazov_14}, etc. Among effects caused by
an oscillating field, the field-induced stabilization of unstable
systems (the dynamical stabilization) should be noted especially
(see, e.g., Ref.~\onlinecite{Bukov_15}). This effect is of
fundamental nature and manifests itself in different areas of
physics
--- from mechanical systems (the Kapitza pendulum~\cite{Kapitza_51}) to
the trapping of particles (the Paul trap~\cite{Paul_90}). Although
the dynamical stabilization has been known for a long time,
related electronic phenomena in nanostructures still await for
detailed analysis. Filling this gap, it was found that an
oscillating field can induce the metastable bound electron states
of various repulsive potentials in nanostructures. Particularly,
the field-mediated electron pairing  appears in nanostructures
containing conduction electrons with different effective masses.
The present article is devoted to the theoretical analysis of this
phenomenon.

The article is organized as follows. In Sec. II, the
one-dimensional (1D) classical dynamics of an electron in the
presence of a repulsive potential and an oscillating field is
considered. In Sec. III, the 1D theory is extended for a quantum
multidimensional case and the field-induced electron pairing is
discussed. The last section contains the conclusion and
acknowledgments.

\section{Model}
To clarify physical origin of the claimed effect, let us consider
the classical dynamics of an electron confined inside a 1D
nanostructure (quantum wire) irradiated by an electromagnetic wave
linearly polarized along the wire. The classical Hamilton function
of the electron in the presence of the repulsive potential $U(x)$
reads
\begin{equation}\label{H}
{\cal H}(x,p)=\frac{1}{2m}\left[p-\frac{e}{c}A(t)\right]^2+U(x),
\end{equation}
where $x$ is the electron coordinate along the quantum wire, $p$
is the generalized momentum of the electron, $e$ is the electron
charge, $m$ is the effective electron mass in the nanostructure
(which can differ from the electron mass in vacuum, $m_e$),
$A(t)=(cE_x/\omega)\cos\omega t$ is the vector potential of the
wave, $E_x$ is the electric field amplitude, and $\omega$ is the
wave frequency. In the particular case of $U(x)=0$, which
corresponds to the free electron driven by the oscillating field
$A(t)$, the electron coordinate oscillates harmonically,
$x(t)=x(0)-x_0\sin\omega t$, where $x_0=eE_x/m\omega^2$ is the
amplitude of the oscillations. To solve the Hamilton problem in
the general case of $U(x)\neq0$, it is convenient to rewrite the
Hamiltonian (\ref{H}) in the reference frame of the oscillating
electron, where its coordinate is $x^\prime=x+x_0\sin\omega t$.
Following the conventional procedure of canonical transformation
of the Hamiltonian (\ref{H}) to the new variables $x^\prime$ and
$p^{\,\prime}$ (see, e.g., Ref.~\onlinecite{Landau_01}), one has
to start from the Lagrangian ${\cal
L}(x,\dot{x})=m\dot{x}^2/2+eA(t)\dot{x}/c-U(x)$ which is
physically equivalent to the Hamiltonian (\ref{H}). Substituting
the electron coordinate $x=x^\prime-x_0\sin\omega t$ into the
Lagrangian, it can be rewritten as ${\cal
L}(x^\prime,\dot{x}^\prime)=m[\dot{x}^\prime-x_0\omega\cos\omega
t]^2/2+eA(t)[\dot{x}^\prime-x_0\omega\cos\omega
t]/c-U(x^\prime-x_0\sin\omega t)$. Therefore, the new generalized
momentum $p^{\,\prime}=\partial{\cal
L}(x^\prime,\dot{x}^\prime)/\partial\dot{x}^\prime$ corresponding
to the new coordinate $x^\prime$ is
$p^{\,\prime}=m\dot{x}^\prime$. As a result, the Hamiltonian
(\ref{H}) written in the new canonical variables $x^\prime$ and
$p^{\,\prime}$ is ${\cal
H}(x^\prime,p^{\,\prime})=p^{\,\prime}\dot{x}^\prime-{\cal
L}(x^\prime,\dot{x}^\prime)={p^{\,\prime\,2}}/{2m}+U(x^\prime-x_0\sin\omega
t)+{m(x_0\omega\cos\omega t)^2}/{2}$. It should be noted that the
last term of this Hamiltonian describes the energy shift arising
from oscillations of the new reference frame. Since this term does
not depend on the canonical variables $x^\prime$ and
$p^{\,\prime}$, it does not effect on dynamics of the electron and
will be omitted in what follows. The kinetic energy of the
electron in the new reference frame, ${p^{\,\prime\,2}}/{2m}$,
does not depend on the oscillating field, whereas the repulsive
potential, $U(x^\prime-x_0\sin\omega t)$, oscillates with the
field frequency $\omega$. Expanding this oscillating potential
into the Fourier series, one can rewrite the Hamiltonian ${\cal
H}(x^\prime,p^{\,\prime})$ as
\begin{equation}\label{H2}
{\cal
H}(x^\prime,p^\prime)=\frac{p^{\,\prime\,2}}{2m}+U_0(x^\prime)+\left[\sum_{n=1}^\infty
U_n(x^\prime)e^{in\omega t}+\mathrm{c.\,c.}\right],
\end{equation}
where
\begin{equation}\label{U}
U_0(x^\prime)=\frac{1}{2\pi}\int_{-\pi}^{\pi}U(x^\prime-x_0\sin\omega
t)d(\omega t),
\end{equation}
is the stationary part of the oscillating potential, and
$U_n(x^\prime)$ are the Fourier series coefficients of the
potential. First of all, let us discuss qualitatively appearance
of the stationary potential (\ref{U}) which is responsible for the
smooth motion of the electron. The most of physically relevant
repulsive potentials
--- including, particularly, the Coulomb repulsion --- are described by a barrier-like function $U(x)$, which
has a maximum at the coordinate $x=0$ and decreases with
increasing $|x|$. Correspondingly, the time-dependent function
$U(x^\prime-x_0\sin\omega t)$ describes the moving potential
barrier which is centered at the oscillating coordinate
$x^\prime=x_0\sin\omega t$. During the period of the oscillations,
this oscillating potential barrier spends most time around the two
points, $x^\prime=\pm x_0$, where the oscillator's velocity,
$x_0\omega\cos\omega t$, is zero [see the green dashed lines in
Fig.~1(a)]. As a consequence, the time-averaged potential
(\ref{U}) acquires a two-barrier structure with a local minimum at
$x^\prime=0$ [see the blue solid line in Fig.~1(a)]. It should be
noted that the local minimum of the integral (\ref{U}) appears for
any integrable barrier-like potential $U(x^\prime-x_0\sin\omega
t)$ if the oscillation amplitude $x_0$ is large enough compared
with the typical scale of the potential. In order to illustrate
this, one can consider the model repulsive potential
$U(x)={e^2}/{\sqrt{a^2+x^2}}$ which can be used, particularly, to
describe the Coulomb interaction of an electron confined in a
quantum wire with the effective thickness $a$. In this case, the
potential~(\ref{U}) reads
\begin{equation}\label{K}
U_0(x^\prime)=\frac{2e^2}{\pi
\sqrt{R(x^\prime)}}\,K\left(\sqrt{\frac{1}{2}-\frac{a^2+x^{\prime2}-x_0^2}
{2R(x^\prime)}}\,\right),
\end{equation}
where $R(x^\prime)=\sqrt{a^2+(x^{\prime}+
x_0)^2}\sqrt{a^2+(x^{\prime}-x_0)^2}$, and $K(\xi)$ is the
complete elliptical integral of the first kind. In accordance with
the aforesaid, the potential (\ref{K}) has a local minimum at
$x^\prime=0$ if the ratio $\lambda=x_0/a$ is greater than the
critical value $\lambda_0\approx2.1$. The local minimum for
$\lambda>\lambda_0$ is clearly seen in Fig.~1(b), where the
potential (\ref{K}) is plotted for different $\lambda$. As a
consequence of the local minimum, the domain of attraction appears
in the core of the repulsive potential.
\begin{figure}[!h]
\includegraphics[width=1.0\linewidth]{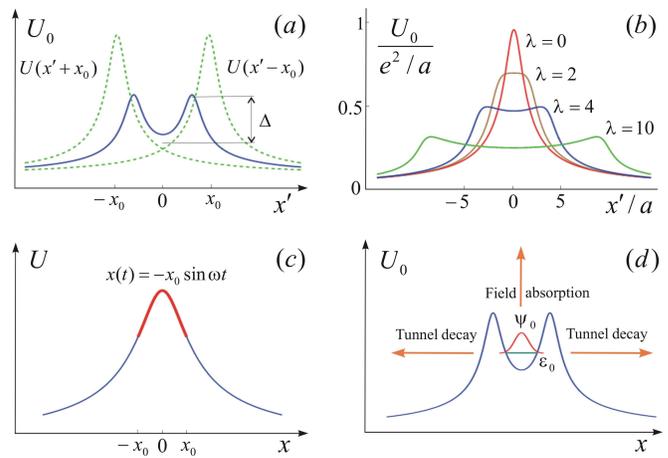}
\caption{The 1D electron problem: (a) scheme of genesis of the
two-barrier potential $U_0(x^\prime)$; (b) the potential (\ref{K})
plotted for different $\lambda=x_0/a$; (c) the electron trajectory
corresponding to the bound state (the heavy red line) confined
near the peak of the repulsive potential $U(x)$; (d) the
metastable bound state with the wave function $\psi_0$ and the
energy $\varepsilon_0$, which can decay due to the tunneling
effect and the field absorption.}
\end{figure}

To complete the analysis of the classical Hamilton problem
(\ref{H}), we have to consider the effect of the oscillating terms
of the Hamiltonian (\ref{H2}) on the electron dynamics in the
domain of attraction. Near the point of local minimum,
$x^\prime=0$, the stationary potential (\ref{U}) can be written as
$U(x^\prime)\approx m\omega^2_0{x}^{\prime 2}/2$, where $\omega_0$
is the eigenfrequency of free electron oscillations near the local
minimum. Then the Hamilton equations corresponding to the
Hamiltonian (\ref{H2}),
$\dot{p}^{\,\prime}=-\partial{\cal{H}}/\partial x^\prime$ and
$\dot{x}^\prime=\partial{\cal{H}}/\partial{p}^{\,\prime}$, lead to
the equation of electron motion near the local minimum:
\begin{equation}\label{de}
\ddot{x}^\prime+\omega_0^2x^\prime=\sum_{n=1}^\infty
[{F_n({x}^\prime)}/{m}]\,e^{in\omega t}+\mathrm{c.\,c.}\,,
\end{equation}
where $F_n(x^\prime)=-\partial U_n(x^\prime)/\partial x^\prime$
are the Fourier components of the periodical force arising from
the oscillating terms of the Hamiltonian (\ref{H2}). Physically,
the dynamic equation (\ref{de}) describes the usual forced
oscillations of an electron under a periodical force with the
harmonics $F_n(x^\prime)$. If the frequency of the force is far
from the resonance, $\omega\gg\omega_0$, the amplitudes of
harmonics of the forced oscillations are substantially smaller
than the typical length of the domain of attraction, $x_0$.
Therefore, the high-frequency periodical force cannot expel the
electron from the local minimum. It should be noted also that the
coordinate dependence of the oscillating terms $F_n(x^\prime)$
leads to the stationary ponderomotive force. However, the
corresponding ponderomotive addition to the stationary potential
(\ref{U}), $\Delta U_0(x^\prime)=\sum_{n=1}^\infty
|F_n(x^\prime)|^2/mn^2\omega^2$, is negligibly small for large
frequencies $\omega$. Thus, the oscillating terms of the
Hamiltonian (\ref{H2}) can be neglected if the frequency of the
driving field, $\omega$, is large enough. In the high-frequency
limit, the dynamics of an electron within the domain of attraction
can be described solely by the stationary potential (\ref{U})
which should be treated as an effective potential renormalized by
the rapidly oscillating field. It follows from the aforesaid that
the local minimum of the potential at $x^\prime=0$ corresponds to
an electron bound at the repulsive potential with the binding
energy $\Delta$ [see Fig.~1(a)]. In the laboratory reference
frame, the bound electron state corresponds to the rapidly
oscillating finite movement of the electron along the stable
trajectory $x(t)=-x_0\sin\omega t$ which is confined near the
potential peak [see Fig.~1(c)].

It follows from the aforesaid that the (meta)stable trajectory
confined in the core of a 1D repulsive potential driven by an
oscillating field (the bound electron state of the repulsive
potential) exists if the oscillation amplitude $x_0$ is large
enough compared with the typical scale of the repulsive potential
$a$. This is why the bound state cannot be described directly
within the established theory of motion in a rapidly oscillating
field, which was elaborated before in the most general form for
small oscillation amplitudes (see, e.g., Sec.~30 of
Ref.~\onlinecite{Landau_01}). As a consequence, the present
analysis of the Hamilton problem (\ref{H})
--- which formally looks like an exercise in classical
mechanics --- contains nontrivial physics under consideration.

\section{Results and discussion}
To go from the classical 1D problem (\ref{H}) to the quantum
multidimensional case, let us start from the Hamiltonian
$\hat{\cal
H}=[\hat{\mathbf{p}}-e\mathbf{A}(t)/c]^2/2m+U(\mathbf{r})$, where
$\hat{\mathbf{p}}=(\hat{p}_x,\hat{p}_y,\hat{p}_z)$ is the momentum
operator, $U(\mathbf{r})$ is the repulsive potential,
$\mathbf{r}=(x,y,z)$ is the radius vector of an electron, and
$\mathbf{A}(t)=(A_x,A_y,A_z)$ is the vector potential of a
homogeneous field which oscillates with the period
$T=2\pi/\omega$. In order to transform the laboratory reference
frame into the rest frame of the oscillating electron, we have to
apply the Kramers-Henneberger unitary
transformation~\cite{Kramers_52,Henneberger_68},
$$\hat{U}(t)=\exp\left\{\frac{i}{\hbar}\int^{\,t}_{-\infty}\left[
\frac{e}{mc}\mathbf{A}(\tau)\hat{\mathbf{p}}-\frac{e^2}{2mc^2}A^2(\tau)
\right]d\tau\right\}. $$ Then the transformed Hamiltonian reads
$\hat{\cal H}^{\prime}= \hat{U}^\dagger(t)\hat{\cal H}\hat{U}(t) -
i\hbar\hat{U}^\dagger(t)\partial_t
\hat{U}(t)=\hat{\mathbf{p}}^2/2m+U(\mathbf{r}-
[{e}/{mc}]\int^{\,t}_{-\infty}\mathbf{A}(\tau)d\tau)$. Expanding
the oscillating potential $U(\mathbf{r},t)$ into a Fourier series,
the transformed Hamiltonian can be written as
\begin{equation}\label{H3}
\hat{\cal
H}^\prime=\frac{\hat{\mathbf{p}}^2}{2m}+U_0(\mathbf{r})+\left[\sum_{n=1}^\infty
U_n(\mathbf{r})e^{in\omega t}+\mathrm{c.\,c.}\right],
\end{equation}
where the stationary part of the potential is
\begin{equation}\label{U0}
U_0(\mathbf{r})=\frac{1}{2\pi}\int_{-\pi}^{\pi}U\left(\mathbf{r}-
\frac{e}{mc}\int^{\,t}_{-\infty}\mathbf{A}(\tau)d\tau\right)d(\omega
t),
\end{equation}
and $U_n(\mathbf{r})$ are the Fourier coefficients of the
oscillating potential. In the theory of laser-atom interaction,
the renormalized potentials of kind (\ref{U0}) are known as the
Kramers-Henneberger (KH)
potentials~\cite{Henneberger_68,Delone_2000}. Certainly, the KH
potential (\ref{U0}) turns into the 1D potential (\ref{U}) if
$\mathbf{r}=(x,0,0)$ and $\mathbf{A}(t)=([cE_x/\omega]\cos\omega
t,0,0)$. Solving the 1D Schr\"odinger problem with the two-barrier
potential $U_0(x)$, one can found the bound state confined between
the two barriers [see Fig.~1(d)]. Within the quantum description,
the bound state is nonstationary and can decay in two ways: (i)
the electron tunneling through the potential barriers; (ii) the
photon absorption arising from the oscillating terms of the
Hamiltonian (\ref{H3}) [the corresponding electron transitions
from the bound state are marked by the arrows in Fig.~1(d)].
However, the increase of the oscillation amplitude $x_0$ increases
both the barrier height and the barrier width [see Fig.~1(b)].
Therefore, the probability of the tunnel decay can be very small
if the oscillation amplitude $x_0$ is large enough. In this case,
the ground bound state confined near the local minimum of the
potential $U_0(x)$ can be described approximately by the
stationary wave function $\psi_0(x)$ with the energy
$\varepsilon_0\sim\hbar\omega_0$ [see Fig.~1(d)]. Regarding the
decay caused by absorption of the field energy, it corresponds to
the electron transitions from the ground bound state with matrix
elements that are very small under the condition
$\omega\gg\omega_0$. As a result, the two mentioned decay
mechanisms can be neglected if both the oscillation amplitude
$x_0=eE_x/m\omega^2$ and the field frequency $\omega$ are large
enough. These two conditions can always be satisfied
simultaneously since one can vary the field strength $E_x$ and the
field frequency $\omega$ independently. Thus, both the classical
description of the problem and the quantum one lead to the same
effect: The metastable bound states of a repulsive potential
driven by an oscillating field appear if the field is both strong
and high-frequency.

Let us extend the consideration for the 2D case corresponding to a
conduction electron confined inside a quantum well irradiated by
an electromagnetic wave. If the quantum well lies in the $(x,y)$
plane and the wave propagates along the $z$ axis, the vector
potential of the wave inside the well can be written as
$\mathbf{A}(t)=([cE_x/\omega]\cos\omega
t,\,[cE_y/\omega]\cos[\omega t-\phi],\,0)$, where $E_{x,y}$ are
amplitudes of the wave along the $x,y$ axes, and $\phi$ is the
wave phase. For definiteness, let us restrict the consideration by
the 2D Coulomb repulsive potential,
$U(\mathbf{r})=e^2/\sqrt{x^2+y^2}$. Substituting this 2D potential
into Eq.~(\ref{U0}), we arrive at the repulsive potential
renormalized by the oscillating field, $U_0(\mathbf{r})$. If the
oscillating field is linearly polarized along the $x$ axis
($E_y=0$), the KH potential (\ref{U0}) takes the form (\ref{K})
with the replacement $a\rightarrow y$. Since this potential has a
two-peak structure without a local minimum, it cannot confine
movement of an electron along the $y$ axis [see Fig.~2(a)]. As a
consequence, the 2D bound states are absent in the case of a
linearly polarized field. However, they can be induced by a
circularly polarized field. Indeed, if the wave is circularly
polarized (the amplitudes are $E_x=E_y=E_0$ and the phase is
$\phi=\pi/2$), the KH potential (\ref{U0}) reads
\begin{equation}\label{U2}
U_0(\rho)=\left\{\begin{array}{rl}
({2e^2}/{\pi\rho_0})K\left({\rho}/{\rho_0}\right),
&{\rho}/{\rho_0}\leq1\\\\
({2e^2}/{\pi\rho})K\left({\rho_0}/{\rho}\right),
&{\rho}/{\rho_0}\geq1
\end{array}\right.,
\end{equation}
where $\rho=\sqrt{x^2+y^2}$ is the polar radius vector, and
$\rho_0=|e|E_0/m\omega^2$ is the radius of the circular trajectory
of a free electron induced by the circularly polarized field (the
oscillation amplitude). Since the potential (\ref{U2}) has a
crater-like structure with a local minimum at $\rho=0$ [see
Fig.~2(b)], it results in the total confinement of the electron
along the $x,y$ axes and, therefore, can produce the bound states.
It should be noted that the bound states of the potential
(\ref{U2}) exist for any oscillation amplitude $\rho_0$, since the
pure 2D Coulomb potential has singularity at $\rho=0$ (i.e., its
typical scale is $a=0$). The renormalized potentials plotted in
Fig.~2 also have weak singularities which look smoothed in the
plots. Namely, Eqs.~(\ref{K}) and (\ref{U2}) involve the
elliptical integral $K(\xi)$, which has the logarithmic
singularity at $\xi=1$. In addition, there are the root
singularities at the potential peaks plotted in Fig.~2(a).
\begin{figure}[!h]
\includegraphics[width=1.0\linewidth]{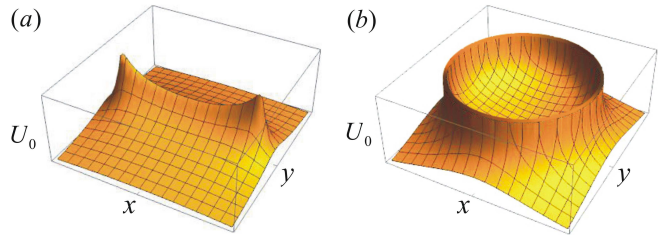}
\caption{The 2D repulsive Coulomb potential renormalized by an
electromagnetic field with different polarizations: (a) linear
polarization along the $x$ axis; (b) circular polarization in the
$x,y$ plane.}
\end{figure}

The one-electron theory developed above can be easily generalized
to describe the interaction of two electrons. In this case, the KH
potential (\ref{U0}) reads
\begin{equation}\label{U00}
U_0(\mathbf{r}_1,\mathbf{r}_2)=\int_{-\pi}^{\pi}U\left(\mathbf{r}_1-\mathbf{r}_2
- \frac{e}{\overline{m}
c}\int^{\,t}_{-\infty}\mathbf{A}(\tau)d\tau \right)\frac{d(\omega
t)}{2\pi},
\end{equation}
where $U(\mathbf{r}_1-\mathbf{r}_2)$ is the initial potential of
repulsive electron-electron interaction,
$\overline{m}=m_1m_2/(m_1-m_2)$ is the reduced electron mass, and
$\mathbf{r}_{j}=(x_{j},y_{j},z_{j})$ and $m_{j}$ with $j=1,2$ are
the radius vectors and effective masses of the interacting
electrons, respectively. It follows from Eq.~(\ref{U00}) that the
oscillating field renormalizes the interaction potential
$U(\mathbf{r}_1-\mathbf{r}_2)$ if the electron masses are not
equal, $m_1\neq m_2$. Otherwise, the field does not change
distance between the electrons and, correspondingly, cannot modify
their interaction. Thus, the field-induced electron-electron
attraction can appear in nanostructures containing conduction
electrons with different effective masses. As an example of such a
nanostructure, let us consider a quantum well consisting of two
layers ($1$ and $2$) filled with a 2D electron gas (2DEG), which
are fabricated using different semiconductor materials and
isolated from each other by a buffer layer with thickness $d$ [see
Fig.~3(a)]. Then the Coulomb interaction of two electrons from the
separated 2D layers $1$ and $2$ can be described by the potential
$U(\mathbf{r}_1-\mathbf{r}_2)=e^2/\sqrt{(x_1-x_2)^2+(y_1-y_2)^2+d^2}$.
If the quantum well is irradiated by a circularly polarized
electromagnetic wave [see Fig.~3(a)], the substitution of this
initial potential into Eq.~(\ref{U00}) yields the renormalized
Coulomb potential
\begin{equation}\label{U3}
U_0(\overline{\rho})=\frac{2e^2}{\pi\sqrt{(\overline{\rho}+\overline{\rho}_0)^2+d^2}}
\,K\left(\sqrt{\frac{4\overline{\rho}_0\overline{\rho}}{(\overline{\rho}+\overline{\rho}_0)^2+d^2}}\,\,\right),
\end{equation}
where $\overline{\rho}=\sqrt{(x_1-x_2)^2+(y_1-y_2)^2}$ and
$\overline{\rho}_0=|e/\overline{m}\,|(E_0/\omega^2)$. The
potential (\ref{U3}) has a local minimum at $\overline{\rho}=0$ if
the ratio $\lambda=\overline{\rho}_0/d$ is greater than the
critical value $\lambda_0\approx1.4$. Physically, this minimum
corresponds to the paired electrons from layers $1$ and $2$ with
binding energy $\Delta$ (see Fig.~3).

The scattering of electrons can destroy the field-induced electron
oscillations which are responsible for the considered effect.
Therefore, the condition $\omega\tau\gg1$ is crucial for the
pairing, where $\tau$ is the mean free time of conduction
electrons. In state-of-the-art semiconductor quantum wells, this
condition can be satisfied for field frequencies of the microwave
range, $\omega\sim10^{11}$~rad/s. Assuming that the reduced mass
is $\overline{m}\sim0.1\,m_e$ and the buffer thickness is
$d\sim$~nm, the electron pair corresponding to the local minimum
of the renormalized potential (\ref{U3}) has the binding energy of
room-temperature scale, $\Delta\sim10^{-2}$~eV, and the typical
size $\overline{\rho}_0\sim10$~nm for irradiation intensity
$I\sim10^{-2}$~W/cm$^2$. Thus, the electron pairing and related
phenomena can be high-temperature for relatively weak irradiation.
\begin{figure}[!h]
\includegraphics[width=1.0\linewidth]{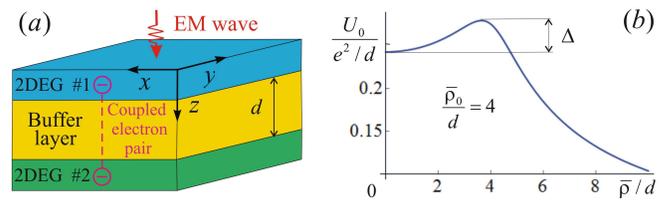}
\caption{The two-layer quantum well irradiated by a circularly
polarized electromagnetic wave: (a) sketch of the system under
consideration; (b) the renormalized Coulomb potential (\ref{U3})
plotted for $\overline{\rho}_0/d=4$.}
\end{figure}

\section{Conclusion}
It is demonstrated that a strong high-frequency electromagnetic
field can induce the metastable bound states of various repulsive
potentials in nanostructures. This leads, particularly, to the
electron pairing in nanostructures containing electrons with
different effective masses. The discussed effect strongly depends
on the field polarization. Namely, a linearly polarized field can
induce electron pairs only in 1D nanostructures, whereas a
circularly polarized field induces them also in 2D nanostructures.
Therefore, semiconductor quantum wells irradiated by a circularly
polarized field look most appropriate for experimental observation
of the pairs. Among a variety of possible effects caused by the
electron pairing, superconductivity mediated by an oscillating
field should be noted especially. To describe this prospective
phenomenon correctly, the solved two-electron problem should be
generalized for the many-electron case. However, such an extension
of the developed theory goes beyond the scope of the present
article and will be done elsewhere.

The work was partially supported by Russian Foundation for Basic
Research (project 17-02-00053) and Ministry of Science and High
Education of Russian Federation (project 3.4573.2017/6.7).

\end{document}